\documentclass[prd,twocolumn,showpacs,linenumbers,groupedaddress,superscriptaddress,amsmath,amssymb]{revtex4}

\usepackage{graphicx}
\usepackage{dcolumn}
\usepackage{bm}
\usepackage{amssymb}
\usepackage{enumerate}
\usepackage{lineno}
\usepackage{multirow} 

\newcommand\st{{\it S}}
\newcommand\vt{{\it V}}
\newcommand\ps{{\rm PS}}
\newcommand\ec{{\rm ECAL}}
\newcommand\hc{{\rm HCAL}}

\newcommand\ee{e^+e^-}

\newcommand\xinv{X\to invisible}

\newcommand\g{\gamma}
\newcommand\mx{m_{X}}

\newcommand\nx{{n}_{X}}

\newcommand\ez{e^- Z \to e^- Z X}
\newcommand\ex{e^- Z \to e^- Z X;~ X \to invisible}
\newcommand\emu{e^- Z \to e^- Z \g; \g \to \mu^+ \mu^-}

\def\address{\@ifstar{\address@star}%
  {\@ifnextchar[{\address@optarg}{\address@noptarg}}}

\usepackage[printwatermark]{xwatermark}
\usepackage{xcolor}
\usepackage{graphicx}
\usepackage{lipsum}

\begin{document}


\title{ 
\begin{center}
 {\Large EUROPEAN LABORATORY FOR PARTICLE PHYSICS}
\end{center}
\vskip1.5cm
\hspace{-3.0cm}{\rightline{\rm  CERN-EP-2021-017}}
\vskip2.5cm
Constraints on New Physics in the Electron  $g-2$ from a Search for Invisible Decays \\  of a Scalar, Pseudoscalar, Vector, and Axial Vector}
\affiliation{\it Universit\"at Bonn, Helmholtz-Institut f\"ur Strahlen-und Kernphysik, 53115 Bonn, Germany} 
\affiliation{\it Joint Institute for Nuclear Research, 141980 Dubna, Russia}
\affiliation{\it Technische Universit\"at M\"unchen, Physik  Department, 85748 Garching, Germany}
\affiliation{\it CERN, European Organization for Nuclear Research, CH-1211 Geneva 23, Switzerland}
\affiliation{\it UCL Departement of Physics and Astronomy, University College London, Gower St. London WC1E 6BT, United Kingdom}
\affiliation{\it Institute for Nuclear Research, 117312 Moscow, Russia}
\affiliation{\it P.N. Lebedev Physical Institute, Moscow, Russia, 119 991 Moscow, Russia}
\affiliation{\it Skobeltsyn Institute of Nuclear Physics, Lomonosov Moscow State University, 119991  Moscow, Russia}
\affiliation{\it Physics Department, University of Patras, 265 04 Patras, Greece} 
\affiliation{\it State Scientific Center of the Russian Federation Institute for High Energy Physics of National Research Center 'Kurchatov Institute' (IHEP), 142281 Protvino, Russia}
\affiliation{\it Departamento de Ciencias F\'{i}sicas, Universidad Andres Bello, Sazi\'{e} 2212, Piso 7, Santiago, Chile}
\affiliation{\it Tomsk Polytechnic University, 634050 Tomsk, Russia}
\affiliation{\it Tomsk State Pedagogical University, 634061 Tomsk, Russia}
\affiliation{\it Universidad T\'{e}cnica Federico Santa Mar\'{i}a, 2390123 Valpara\'{i}so, Chile}
\affiliation{\it ETH Z\"urich, Institute for Particle Physics and Astrophysics, CH-8093 Z\"urich, Switzerland}
\affiliation{\it SAPHIR Millennium Institute of ANID, Chile}
\author{Yu.~M.~Andreev}\affiliation{\it Institute for Nuclear Research, 117312 Moscow, Russia}
\author{D.~Banerjee}\affiliation{\it CERN, European Organization for Nuclear Research, CH-1211 Geneva 23, Switzerland}
\author{J.~Bernhard}\affiliation{\it CERN, European Organization for Nuclear Research, CH-1211 Geneva 23, Switzerland}
\author{V.~E.~Burtsev}\affiliation{\it Joint Institute for Nuclear Research, 141980 Dubna, Russia}
\author{A.~G.~Chumakov}\affiliation{\it Tomsk Polytechnic University, 634050 Tomsk, Russia}\affiliation{\it Tomsk State Pedagogical University, 634061 Tomsk, Russia}
\author{D.~Cooke}\affiliation{\it UCL Departement of Physics and Astronomy, University College London, Gower St. London WC1E 6BT, United Kingdom}
\author{P.~Crivelli}\affiliation{\it ETH Z\"urich, Institute for Particle Physics and Astrophysics, CH-8093 Z\"urich, Switzerland}
\author{E.~Depero}\affiliation{\it ETH Z\"urich, Institute for Particle Physics and Astrophysics, CH-8093 Z\"urich, Switzerland}
\author{A.~V.~Dermenev}\affiliation{\it Institute for Nuclear Research, 117312 Moscow, Russia}
\author{S.~V.~Donskov}\affiliation{\it State Scientific Center of the Russian Federation Institute for High Energy Physics of National Research Center 'Kurchatov Institute' (IHEP), 142281 Protvino, Russia}
\author{R.~R.~Dusaev}\affiliation{\it Tomsk Polytechnic University, 634050 Tomsk, Russia}
\author{T.~Enik}\affiliation{\it  Joint Institute for Nuclear Research, 141980 Dubna, Russia}
\author{N.~Charitonidis}\affiliation{\it CERN, European Organization for Nuclear Research, CH-1211 Geneva 23, Switzerland}
\author{A.~Feshchenko}\affiliation{\it  Joint Institute for Nuclear Research, 141980 Dubna, Russia}
\author{V.~N.~Frolov}\affiliation{\it  Joint Institute for Nuclear Research, 141980 Dubna, Russia}
\author{A.~Gardikiotis}\affiliation{\it Physics Department, University of Patras, 265 04 Patras, Greece}
\author{S.~G.~Gerassimov }\affiliation{\it Technische Universit\"at M\"unchen, Physik  Department, 85748 Garching, Germany}\affiliation{\it P.N. Lebedev Physical Institute, Moscow, Russia, 119 991 Moscow, Russia}
\author{S.~N.~Gninenko\footnote[1]{Corresponding author: Sergei.Gninenko@cern.ch}}
\affiliation{\it Institute for Nuclear Research, 117312 Moscow, Russia}
\author{M.~H\"osgen}\affiliation{\it Universit\"at Bonn, Helmholtz-Institut f\"ur Strahlen-und Kernphysik, 53115 Bonn, Germany}
\author{V.~A.~Kachanov}\affiliation{\it State Scientific Center of the Russian Federation Institute for High Energy Physics of National Research Center 'Kurchatov Institute' (IHEP), 142281 Protvino, Russia}
\author{A.~E.~Karneyeu}\affiliation{\it Institute for Nuclear Research, 117312 Moscow, Russia}
\author{G.~Kekelidze}\affiliation{\it  Joint Institute for Nuclear Research, 141980 Dubna, Russia}
\author{B.~Ketzer}\affiliation{\it Universit\"at Bonn, Helmholtz-Institut f\"ur Strahlen-und Kernphysik, 53115 Bonn, Germany}
\author{D.~V.~Kirpichnikov}\affiliation{\it Institute for Nuclear Research, 117312 Moscow, Russia}
\author{M.~M.~Kirsanov}\affiliation{\it Institute for Nuclear Research, 117312 Moscow, Russia}
\author{V.~N.~Kolosov}\affiliation{\it State Scientific Center of the Russian Federation Institute for High Energy Physics of National Research Center 'Kurchatov Institute' (IHEP), 142281 Protvino, Russia}
\author{I.~V.~Konorov}\affiliation{\it Technische Universit\"at M\"unchen, Physik  Department, 85748 Garching, Germany}\affiliation{\it P.N. Lebedev Physical Institute, Moscow, Russia, 119 991 Moscow, Russia} 
\author{S.~G.~Kovalenko}\affiliation{\it Departamento de Ciencias F\'{i}sicas, Universidad Andres Bello, Sazi\'{e} 2212, Piso 7, Santiago, Chile}
\author{V.~A.~Kramarenko}\affiliation{\it  Joint Institute for Nuclear Research, 141980 Dubna, Russia}\affiliation{\it Skobeltsyn Institute of Nuclear Physics, Lomonosov Moscow State University, 119991  Moscow, Russia}
\author{L.~V.~Kravchuk}\affiliation{\it Institute for Nuclear Research, 117312 Moscow, Russia}
\author{ N.~V.~Krasnikov}\affiliation{\it  Joint Institute for Nuclear Research, 141980 Dubna, Russia}\affiliation{\it Institute for Nuclear Research, 117312 Moscow, Russia}
\author{S.~V.~Kuleshov}\affiliation{\it Departamento de Ciencias F\'{i}sicas, Universidad Andres Bello, Sazi\'{e} 2212, Piso 7, Santiago, Chile}\affiliation{\it SAPHIR Millennium Institute of ANID, Chile}
\author{V.~E.~Lyubovitskij}\affiliation{\it Tomsk Polytechnic University, 634050 Tomsk, Russia}\affiliation{\it Tomsk State Pedagogical University, 634061 Tomsk, Russia}\affiliation{\it Universidad T\'{e}cnica Federico Santa Mar\'{i}a, 2390123 Valpara\'{i}so, Chile}
\author{V.~Lysan}\affiliation{\it  Joint Institute for Nuclear Research, 141980 Dubna, Russia}
\author{V.~A.~Matveev}\affiliation{\it  Joint Institute for Nuclear Research, 141980 Dubna, Russia}
\author{Yu.~V.~Mikhailov}\affiliation{\it State Scientific Center of the Russian Federation Institute for High Energy Physics of National Research Center 'Kurchatov Institute' (IHEP), 142281 Protvino, Russia}
\author{L.~Molina Bueno}\affiliation{\it ETH Z\"urich, Institute for Particle Physics and Astrophysics, CH-8093 Z\"urich, Switzerland}
\author{D.~V.~Peshekhonov}\affiliation{\it  Joint Institute for Nuclear Research, 141980 Dubna, Russia}
\author{V.~A.~Polyakov}\affiliation{\it State Scientific Center of the Russian Federation Institute for High Energy Physics of National Research Center 'Kurchatov Institute' (IHEP), 142281 Protvino, Russia}
\author{B.~Radics}\affiliation{\it ETH Z\"urich, Institute for Particle Physics and Astrophysics, CH-8093 Z\"urich, Switzerland}
\author{R.~Rojas}\affiliation{\it Universidad T\'{e}cnica Federico Santa Mar\'{i}a, 2390123 Valpara\'{i}so, Chile}
\author{A.~Rubbia}\affiliation{\it ETH Z\"urich, Institute for Particle Physics and Astrophysics, CH-8093 Z\"urich, Switzerland}
\author{V.~D.~Samoylenko}\affiliation{\it State Scientific Center of the Russian Federation Institute for High Energy Physics of National Research Center 'Kurchatov Institute' (IHEP), 142281 Protvino, Russia}
\author{H.~Sieber}\affiliation{\it ETH Z\"urich, Institute for Particle Physics and Astrophysics, CH-8093 Z\"urich, Switzerland}
\author{D.~Shchukin}\affiliation{\it P.N. Lebedev Physical Institute, Moscow, Russia, 119 991 Moscow, Russia}
\author{V.~O.~Tikhomirov}\affiliation{\it P.N. Lebedev Physical Institute, Moscow, Russia, 119 991 Moscow, Russia}
\author{I.~Tlisova}\affiliation{\it Institute for Nuclear Research, 117312 Moscow, Russia} 
\author{A.~N.~Toropin}\affiliation{\it Institute for Nuclear Research, 117312 Moscow, Russia}
\author{A.~Yu.~Trifonov}\affiliation{\it Tomsk Polytechnic University, 634050 Tomsk, Russia}\affiliation{\it Tomsk State Pedagogical University, 634061 Tomsk, Russia}
\author{B.~I.~Vasilishin}\affiliation{\it Tomsk Polytechnic  University, 634050 Tomsk, Russia}
\author{P.~V.~Volkov}\affiliation{\it  Joint Institute for Nuclear Research, 141980 Dubna, Russia}\affiliation{\it Skobeltsyn Institute of Nuclear Physics, Lomonosov Moscow State University, 119991  Moscow, Russia}
\author{V.~Yu.~Volkov}\affiliation{\it Skobeltsyn Institute of Nuclear Physics, Lomonosov Moscow State University, 119991  Moscow, Russia}

%
%
\collaboration{The NA64 Collaboration}\noaffiliation
\vskip 0.25cm

\date{\today}

\begin{abstract}
We performed a search for  a new generic $X$ boson, which could be a scalar ($S$),  pseudoscalar ($P$), vector ($V$) or an axial vector ($A$) particle  produced in the 100 GeV electron scattering off nuclei, $\ez$,  followed by its  invisible decay in the NA64 experiment at  CERN.  No evidence for such process was found in the full NA64 data set  of $2.84\times 10^{11}$ electrons on target.  We place new  bounds on the $S,P,V, A$  coupling strengths to electrons, and set constraints  on their contributions  to the electron anomalous magnetic moment $a_e$,  
 $|\Delta a_{X}| \lesssim 10^{-15} - 10^{-13}$ for the $X$ mass region $m_X\lesssim 1$ GeV.  These results are  an order of magnitude more sensitive compared to the current accuracy 
 on $a_e$ from  the electron $g-2$ experiments  and recent high-precision determination of the fine structure constant.
\end{abstract}

\maketitle

Searching for new physics (NP)  with mass below the electroweak scale ($\ll 100$ GeV) at the high-intensity and high-precision  frontiers has received significant attention in recent years 
\cite{jr,Essig:2013lka,report1,report2,pbc-bsm, esu, berlin, gaia}. 
 Motivations for searches of  feebly-coupled particles in the low-mass range  come from  the evidence for NP in the neutrino and dark matter sectors, and are well supported by theoretical arguments, see,  e.g. Refs.\cite{jr,berlin,gaia, bf,pf,prv,mp,arkani}. Existing anomalies  observed  in particle experiments also contribute to the field.    Well-known examples are the current 
 muon $g-2$ anomaly - the $\simeq 3.6\sigma$ discrepancy between the predicted and observed value of the muon anomalous magnetic moment \cite{g-2_mu},  or the X17 anomaly  - 
 an excess of $\ee$ events in the $^8$Be and $^4$He nuclei transitions \cite{atomki1,atomki2}, which  might be  explained by NP models at low-mass scale, 
 see, e.g. Refs.\cite{gk,feng}. These anomalies are being scrutinized in the upcoming experiments at Fermilab and JPARC \cite{fnal,jpark}, and  with NA64 at  CERN \cite{na64be1,na64be2,na64be3},  respectively. 
\par Recently, a new puzzle indicating the possible presence of NP in the electron $g-2$ has emerged.  The precise measurements performed at Laboratoire Kastler Brossel (LKB) with $^{87}$Rb rubidium atoms report a new value for the fine-structure constant $\alpha^{-1}$~=~137.035999206(11) with a relative accuracy of 81 parts per trillion \cite{lkb}. This result improves the accuracy on $\alpha$  by 2.5 over the previous measurements performed at  Berkeley  with $^{137}$Cs atoms \cite{berkeley} but, surprisingly, it reveals a $5.4 \sigma$ difference from this latest result.
 Using these measurements of the fine-structure constant, the Standard Model (SM)  prediction of the anomalous magnetic moment of the electron,  $a _e= (g-2)_e/2$ 
 \cite{aoyama1, aoyama2}, is $1.6~ \sigma$  lower  and  $-2.4 ~\sigma$  higher than the direct experimental measurement of $a_e^{exp}$ \cite{gabri}:    
\begin{eqnarray}
&&\Delta a_e= a^{exp}_{e} - a^{ LKB}_e = (4.8 \pm 3.0) \times 10^{-13} \label{eq:lkb} \\
&&\Delta a_e= a^{exp}_{e} - a^{ B}_e = (-8.8 \pm 3.6) \times 10^{-13} \label{eq:berk}
\end{eqnarray}
for the LKB and Berkeley measurements, respectively. The errors  on $\Delta a_e$ are  dominated mostly by the  uncertainty in $a^{exp}_{e}$.
As the SM predicts a certain value of the $a_{e}$  \cite{aoyama1, aoyama2} the measurements of this parameter in different processes should be consistent with each other. 
With new measurements and improved SM calculations, one hopes to clarify whether the deviations of Eqs.(\ref{eq:lkb},\ref{eq:berk}) are a result  of yet unknown experimental  errors, or 
  it is  a sign of new  physics in the electron $g-2$ \cite{giud}.  This  motivates  recent  significant efforts towards  
  possible  explanation of the  deviation, in particular the discrepancy of Eq.(\ref{eq:berk}), with a NP effect, see, e.g., Refs.\cite{nvk}-\cite{gkm}. 
\par In this Letter,  we study the question of whether a new light $X$ boson could  contribute  to the electron $g-2$. We consider models with a generic  $X$ in sub-GeV mass range,  which could be a  scalar ($S$), pseudoscalar ($P$), vector ($V$), or an axial~vector ($A$) particle feebly  coupled  to electrons. It is assumed that the  $X$ decays predominantly invisibly, $\Gamma (X\rightarrow invisible)/\Gamma_{tot} \simeq 1$, e.g. into dark sector particles, thus escaping  stringent constraints placed today on the visible   decay modes of the $X$ into SM particles from collider, fixed-target, and atomic experiments  \cite{pdg}.  The most stringent limits on the invisible  $X$ in the sub-GeV mass range are obtained, so far, for the $V$ case of dark photons coupled to electrons through the mixing with the ordinary photons by the NA64 \cite{na64prl19} and $BABAR$ \cite{babarg-2} experiments, leaving a large area of the parameter space for the generic $X$  still unexplored.
 Various aspects of  such invisible  $X$  weakly coupled to leptons including  possible phenomenological implications can be found in 
Refs.\cite{jr,Essig:2013lka,report1,report2,pbc-bsm, esu, berlin, gaia, gkm,Gninenko:2018ter,Kirpichnikov:2020tcf}. 
\par The $e-X$-interaction with   the  coupling strength  $g_X$ defined as $g_X =\varepsilon_X e$ (here $\varepsilon_X$ is a parameter and $e$ is the charge of the electron) is given for the $S,P,V, A$ cases  by phenomenological Lagrangians:
\begin{eqnarray}
\mathcal{L}_S &=& g_S \overline{e} e S \nonumber \\
\mathcal{L}_P &=& ig_P \overline{e}\gamma_5 e P \nonumber \\
\mathcal{L}_V &=& g_V \overline{e}\gamma_\mu e V_\mu \nonumber \\
\mathcal{L}_A &=& g_A \overline{e}\gamma_\mu \gamma_5 e A_\mu  
\label{eq:Lagrangian}
\end{eqnarray}
\begin{figure}[tbh!!]
\includegraphics[width=.5\textwidth]{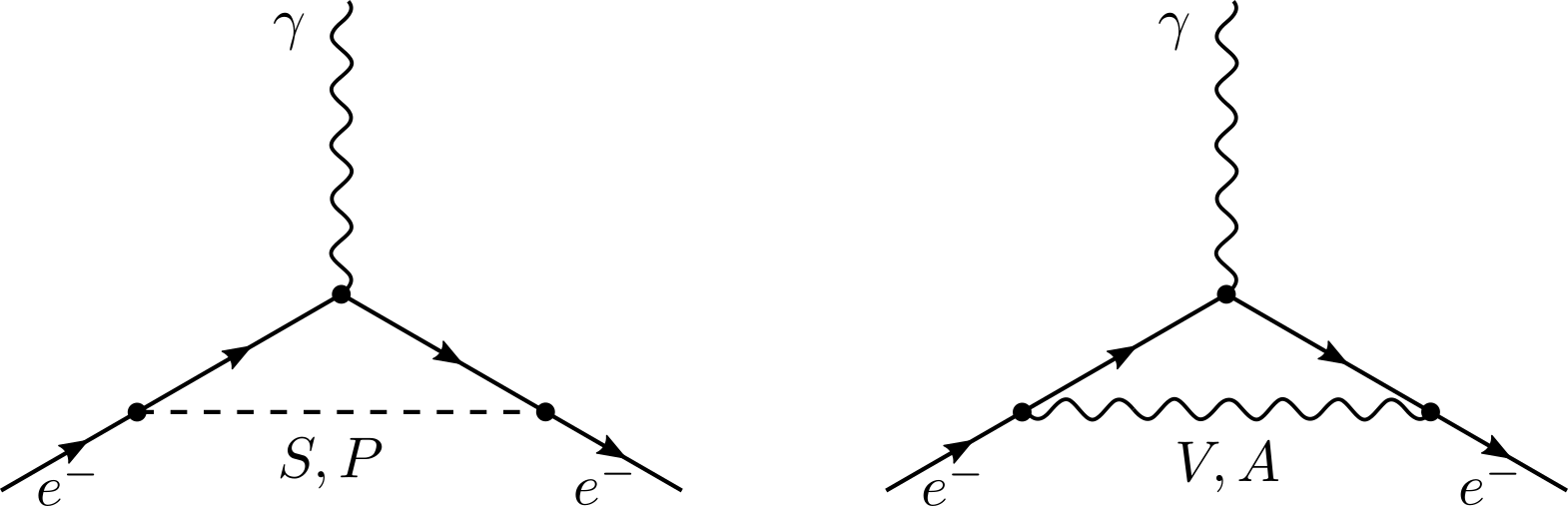}%
\vskip-0.cm{\caption{One-loop contribution of the $S$ and $P$ (left panel) and the $V$ and $A$ (right panel) to $\Delta a_e$. \label{fig:diagr}}}
\end{figure}
The corresponding  one-loop contributions
to the $(g-2)_e$ factor induced by diagrams shown in Fig. \ref{fig:diagr} are given by:
\begin{eqnarray}
\Delta a_S &=& \frac{g_S^2}{4\pi^2} \bigl(\frac{m_e}{m_X}\bigr)^2\bigl[{\rm ln}\frac{m_X}{m_e} -\frac{7}{12}\bigr] \label{eq:contribS} \\
\Delta a_P &=& \frac{g_P^2}{4\pi^2} \bigl(\frac{m_e}{m_X}\bigr)^2\bigl[- {\rm ln}\frac{m_X}{m_e} +\frac{11}{12}\bigr] \label{eq:contribP} \\
\Delta a_V &=& \frac{g_V^2}{4\pi^2} \bigl(\frac{m_e}{m_X}\bigr)^2 \frac{1}{3} \label{eq:contribV} \\
\Delta a_A &=& \frac{g_A^2}{4\pi^2} \bigl(\frac{m_e}{m_X}\bigr)^2 \bigl(-\frac{5}{3} \bigr)  \label{eq:contribA}
\label{eq:contribA*}
\end{eqnarray}
assuming that $m_X \gg m_e$. One can see that presumably a scalar and a vector can explain the positive deviation of Eq.(\eqref{eq:lkb}), while only a  pseudoscalar and an axial vector  could explain the negative value of Eq.(\eqref{eq:berk}). The required couplings $g_X$ to explain deviations of Eqs.(\ref{eq:lkb},\ref{eq:berk}) are in the range $10^{-3} \lesssim |g_X| \lesssim 10^{-4}$ which  is accessible to the NA64 search, thus making it interesting.
\par  The  method of the search, discussed in this work and proposed in Refs.~\cite{Gninenko:2013rka,Andreas:2013lya}, 
 is based on the detection of the missing energy, carried away by the hard bremsstrahlung $X$
 produced in the process $\ex$ of high-energy electrons scattering in an active beam dump.  
 The NA64  experiment  employed  a  100 GeV pure electron beam, using the H4 beam-line of the CERN's North Area. The beam was slowly extracted towards NA64 in 4.8 s spills, and had an intensity up to  $\simeq 10^7$ electrons per spill.
 The e$^-$ beam  was defined by the  scintillator ($\st$) and veto ($\vt_1$) counters.  A magnetic spectrometer  consisting of two successive  dipole magnets with the integral  magnetic strength of $\simeq$7 T$\cdot$m  and a low-material-budget tracker consisting of  a set of Micromegas (MM),  Straw-Tube (ST) and Gaseous Electron Multiplier (GEM) chambers allowed to measure the incoming  $e^-$ momenta with the precision $\delta p/p \simeq 1\%$ \cite{Banerjee:2015eno}. 
The synchrotron radiation (SR) emitted in the magnets   was used for the  electron identification and their efficient  tagging with a SR detector (SRD)\cite{na64srd}, which was an array  of a  Pb-Sc sandwich calorimeter of a fine segmentation. By using the SRD the intrinsic hadron contamination of the beam of the order of $\sim 1\%$ was further suppressed to a negligible level.
The downstream part of the detector was equipped with an electromagnetic ({\it e-m}) calorimeter (ECAL),  a  matrix of $6\times 6 $  Shashlik-type modules   assembled from  Pb and Sc plates  serving as an active beam-dump target  for  measurement of the electron energy  $E_{\ec} $. 
 Each ECAL module has $\simeq 40$ radiation  lengths ($X_0$) with the first 4$X_0$ serving as a preshower detector (PS).   
 Further downstream the detector was equipped with a  high-efficiency veto counter ($\vt_2$), and a hermetic hadronic calorimeter (HCAL) of $\simeq 30$ nuclear interaction lengths in total. The  HCAL  was  used as an efficient veto against hadronic secondaries  and also  to detect muons produced in $e^-$ interactions  in the target.
\par The search described in this paper uses the data samples of $n_{EOT}=2.84\times 10^{11}$ electrons on target (EOT),  collected  in  the years 2016, 2017 and 2018 (runs I,II, and III,  respectively) at the beam intensities mostly  in the range  $\simeq(5-9)\times 10^6$   e$^-$ per spill   with the hardware trigger  \cite{na64prl17,na64prd18,na64prl19}
\begin{equation}
Tr(X) = \Pi \st_i \cdot \vt_1 \cdot \ps(>E^{th}_{\ps}) \cdot \overline{\ec}(< E^{th}_{\ec}), 
\label{trigger}
\end{equation}
accepting events with  in-time hits in  beam-defining counters $S_i$ and clusters in the PS and ECAL with the energy exceeding the thresholds  $ E^{th}_{\ps}\simeq 0.3$ GeV and $E^{th}_{\ec} \lesssim 80$ GeV, respectively.  
The missing energy  events have the signature 
\begin{equation}
S(X) = Tr(X) \cdot  {\rm Track} (P_e) \cdot \vt_2 (< E^{th}_{\vt_2}) \cdot \hc(< E^{th}_{\hc})
\label{sign}
\end{equation}
with the incoming track momentum $P_e\simeq 100\pm 3$ GeV, and $V_2$  and HCAL zero-energy  deposition, defined as energy  below the thresholds 
 $ E^{th}_{\vt_2} \simeq 1$ MIP (minimum ionizing particle) and  $ E^{th}_{\hc} \simeq 1$ GeV, respectively.
 Data from these three runs, were processed with  selection criteria similar to the one used in Refs. \cite{na64prd18,na64prl19} and finally analysed as described below. 
\par
\begin{figure}[tbh!!]
\includegraphics[width=.45\textwidth]{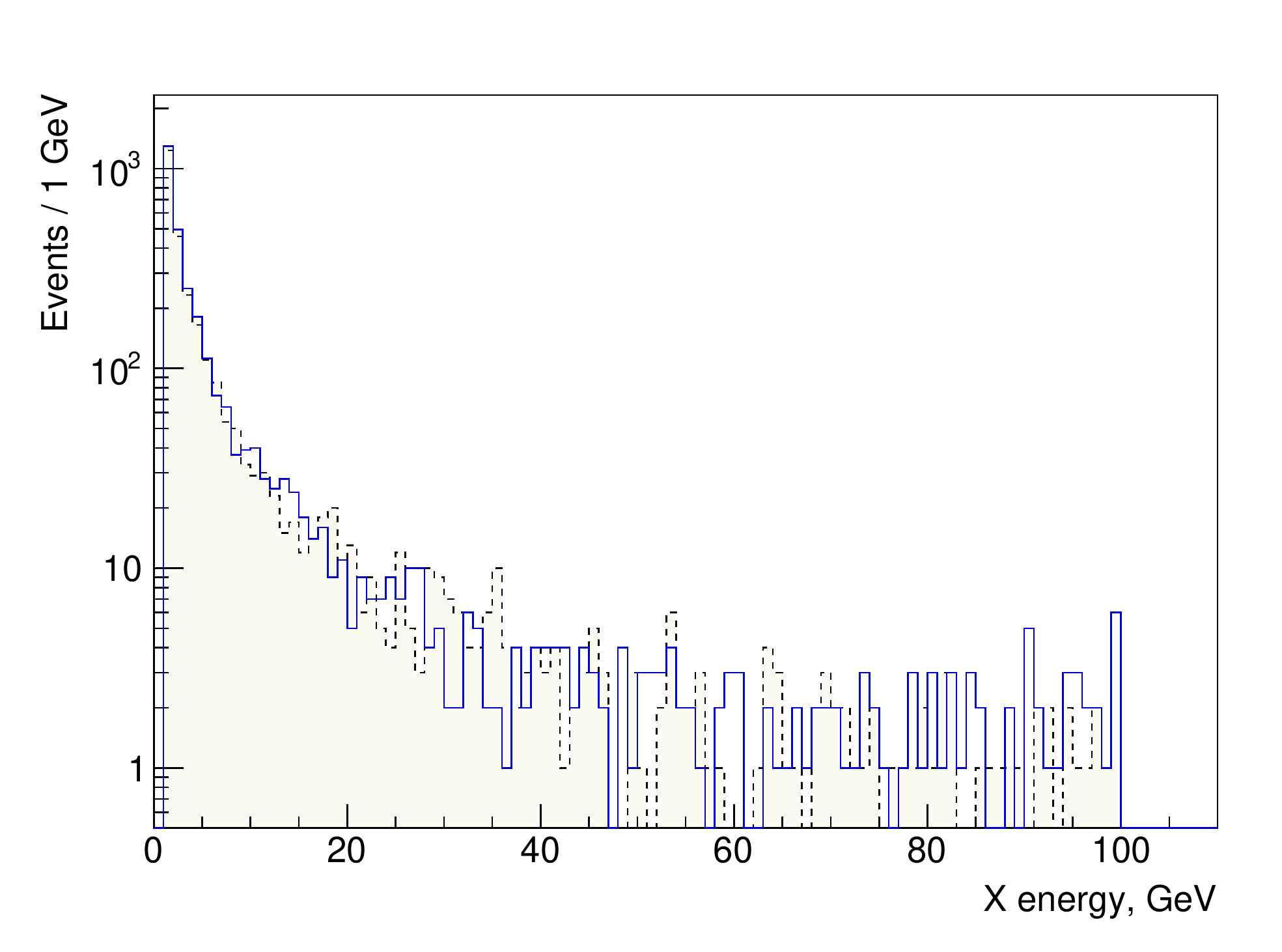}%
\vskip-0.cm{\caption{The emission spectra of the  20 MeV {\it P}(solid line) and {\it V}(dashed line) particles produced from the interactions of the 100 GeV electron beam  in the ECAL  target  obtained 
from the ETL calculations. The spectra are normalized to the same number of EOT.  \label{ecal-spectra}}}. 
\end{figure}

  A detailed  Geant4 \cite{Agostinelli:2002hh, geant} based Monte Carlo (MC) simulation  was used to study detector performance and signal acceptance, to simulate backgrounds and selection cuts.  For the calculations of the signal yield we used the fully Geant4  compatible package DMG4 \cite{dmg4}.  Using this package the  production of $X$ in the process $\ex$
  has been simulated for each type of interactions listed in  Eq.(\ref{eq:Lagrangian}) with  cross-sections obtained from exact tree-level (ETL) calculations, see, e.g.,   Refs. \cite{gkkk,gkkketl,dkk}. The produced signal samples were processed by the same reconstruction program as the real data and passed through the same selection criteria.
The total number  $\nx$ of the  produced $X$ per single electron on target (EOT) was calculated  as 
\begin{equation}
\nx(g_X,~m_{X},~E_0)   =  \frac{\rho N_{A}}{A_{{\rm Pb}}}   \sum_i   n(E_0,E_e,s) \sigma_{X}(E_{e}) \Delta s_i
\label{AprYields}
\end{equation} 
where  $ \rho$ is density of the target, $N_A$ is the 
Avogadro's number, $A_{{\rm Pb}}$ is the  
Pb atomic mass, $n(E_0,E_e,s)$ is  the number of $e^\pm$ in the {\it e-m} shower at the depth $s$ (in radiation lengths) with the energy $E_e$  within the target of total thickness $T$, and 
$\sigma(E_e)$ is the cross section of the $X$ production in the kinematically allowed region up to 
$E_{X}\simeq E_e$ by an electron with the energy $E_e$ in the  reaction $\ex$. The latter  depends in particular on the coupling and mass  $g_X, ~m_{X}$, and the beam energy $E_0$.
The $X$ energy distribution $\frac{d n_{X}}{d E_{X}}$ was  calculated for each case by taking into account the  corresponding differential cross-section $\frac{d\sigma(E_e, E_{X})}{dE_{X}}$, as described in Ref.\cite{gkkketl}. An example of the simulated {\it X} (or missing) energy spectrum  in the  target calculated by using the detailed simulation of {\it e-m} shower development  by Geant4 is shown for  the $P$ and  $V$ cases in Fig.~\ref{ecal-spectra} for the mass $\mx = 20$ MeV. 
The expected number of $X$ events  in our detector from the reaction $\ex$ was determined for each $X$ interaction type also by comparison  to the  rare process of dimuon production, $\emu$,  which has  a well-known reaction rate. These events originate from the QED reaction in the ECAL,  dominated by the  hard bremsstrahlung photon conversion into dimuon pairs on a target nucleus and accompanied by small energy deposition in the HCAL, thus   mimicking   the $\xinv$ decay events below the two-MIP threshold. The reaction
  was also used  as a benchmark process  allowing us  to verify the reliability of the MC simulation, correct  the signal acceptance,  cross-check  systematic uncertainties  and background estimate \cite{na64prd18, na64prl19}. Good agreement was found between the observations and simulations. Using rare dimuon events as a crosscheck for normalization to the signal modes cancels many systematic uncertainties by keeping selection cuts identical whenever possible.
\par In order to avoid biases in the determination of the  selection criteria for signal events, a blind analysis similar to the one described in 
Ref.\cite{na64prl19} was performed.  The signal box ($E_{\ec} < 50~{\text GeV }; E_{\hc} < 1~{\text GeV }$) was defined based on the energy spectrum calculations for $X$s emitted by $e^\pm$ from the {\it e-m} shower generated by the primary  $e^-$s in the  ECAL \cite{gkkk, gkkketl}  and the HCAL zero-energy threshold determined mostly by the noise of the read-out electronics. 
  Finally,  to maximize the acceptance for  signal events  and to minimize  backgrounds we used  the following   selection criteria: 
(i) The incoming electron track momentum  should be within $100\pm 3$ GeV; 
(ii)  The SRD energy should be within the range of the SR energy emitted by $e^-$s in the magnets and in time with the trigger;
(iii)  The shower shape  in the ECAL should be  consistent with the  one expected for the signal shower \cite{gkkk};
(iv) There should be only a single track activity in the tracker chambers upstream of the dump in order to reject  interactions in the beam line  materials, and  no activity in $\vt_2$. 
\par The dominant background for $\ex$ arises from the interactions of the $e^-$ beam  in the downstream part of the detector resulting in hadron electro-production  
in the beam line materials.  In rare cases, these reactions are accompanied by the emission of  large-angle (high $p_T$) hadronic secondaries faking  the  signal due to the insufficient downstream detector coverage. Charged secondaries were rejected by requiring  no additional tracks or hits in the downstream ST chambers, which have the largest transverse acceptance in our setup.  We also requested  no extra in-time hits  upstream of the magnets and at most one extra 
in-time hit downstream of the magnets in the MM chambers. The remaining background from the large-angle neutral hadronic secondaries was evaluated  mainly from data 
by the extrapolation of events  from the sideband  ($E_{\ec} > 50~{\text GeV }; E_{\hc} < 1~{\text GeV }$) into the signal region and  assessing the systematic errors by varying the fit functions selected as described in Ref. \cite{na64prd18}. The shape of the extrapolation functions  was evaluated  from the study of a larger data sample of events from 
 hadronic  $e^-$ interactions in the dump, which was also cross-checked with  simulations.  Another background from punch-through of leading (with energy $\gtrsim 0.5~E_0$) neutral hadrons $(n, K^0_L)$ produced in the $e^-$  interactions in the target, was studied  by using  events from the region ($E_{\ec} < 50~{\text GeV }; E_{\hc} > 1~{\text GeV }$),  which were pure neutral hadronic secondaries produced in the ECAL. Its level was estimated from the data  by using  the longitudinal segmentation of the HCAL and  the   punch-through probability estimated conservatively
and was found to be negligible. Several other background sources that may fake the  signal, such as loss of dimuons due 
to statistical fluctuations of the signal or muon decays,  and  decays in flight of mistakenly SRD tagged  beam
$\pi$, $K$  were simulated with the full statistics of the data and also were found to be negligible.   
   After determining all the selection criteria and background levels, we unblinded the signal region and found 0 events  consistent with  $0.53\pm 0.17$ events from the 
   conservative background estimations \cite{na64prl19} allowing  us to obtain the $m_{X}$-dependent upper limits on the $e-X$ coupling strengths. 
\begin{figure}[tbh!!]
\includegraphics[width=0.5\textwidth]{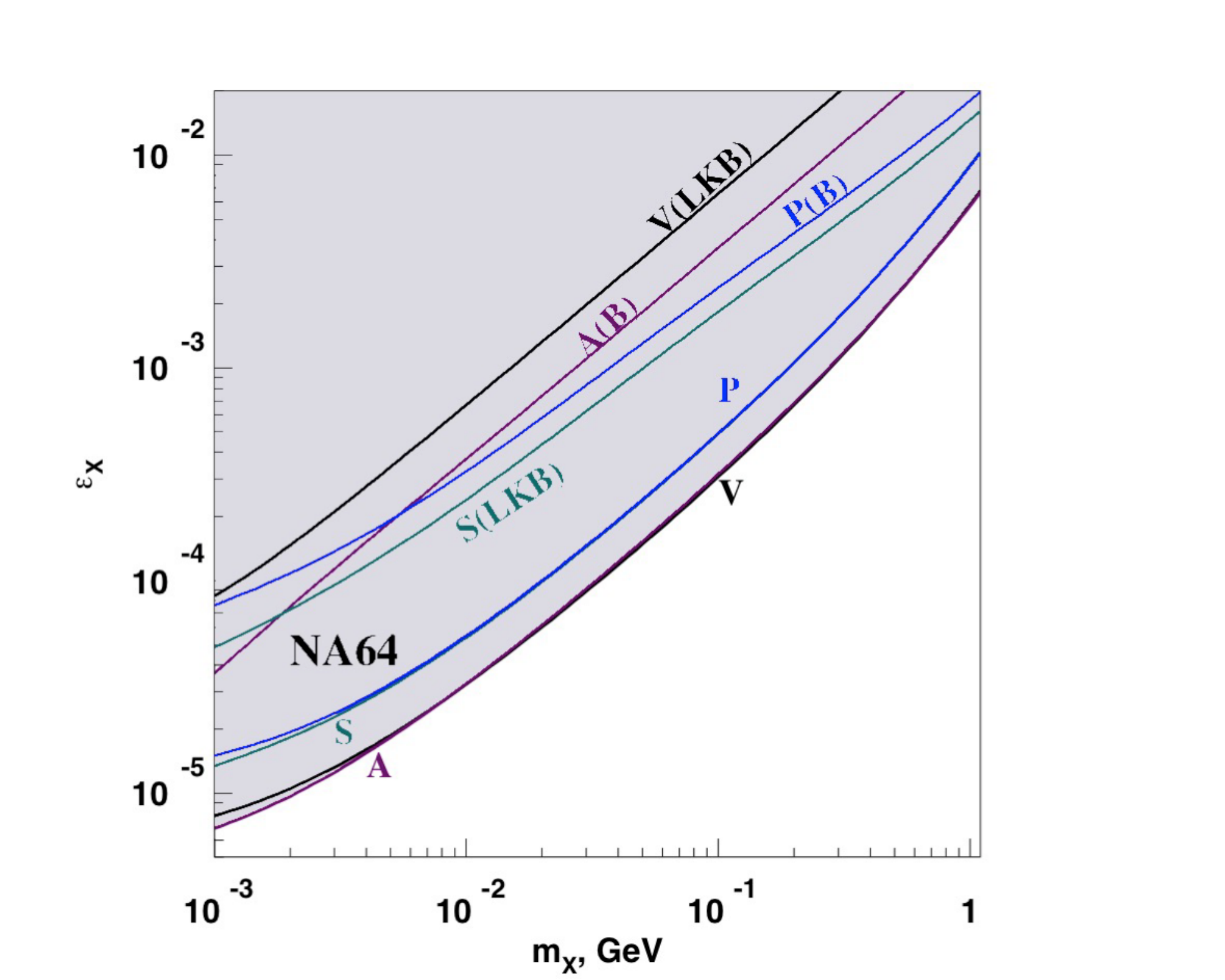}
\caption{The   90\% C.L. upper limits  on the coupling parameter $\varepsilon_X$  in the ($m_{X}, \varepsilon_X $) plane obtained by NA64  and presented in comparison with the
bounds derived from  the results of the LKB \cite{lkb} and Berkeley (B) \cite{berkeley} experiments. The limits are  shown by lines labeled with the $X$  type of the same color. 
} \label{fig:limits}
\end{figure}     
 \par  The overall signal efficiency   $\epsilon_{X}$ defined as the 
product of signal efficiencies accounting for the  geometrical acceptance, the track, SRD, $\vt_2$  and HCAL reconstruction,  and  the DAQ dead time 
was found  to be slightly dependent on $m_{X},E_{X}$  values \cite{na64prl19}.  The signal-event reconstruction efficiency $\epsilon_{\ec}$ was estimated  as a function of  energy deposited in the ECAL
for different $X$ masses.  Compared to the ordinary {\it e-m} shower,  the $\epsilon_{\ec}$  value  for  a shower from  $X$ event 
 has to be corrected due to difference in the {\it e-m} showers development at  the early stage in the ECAL PS  \cite{gkkk}. 
 Depending on the  energy threshold  in the PS ($E^{th}_{\ps}$) used in trigger  \eqref{trigger} this correction   was  
$\lesssim (5\pm 3)\%$ dominated by  the errors due to the $E^{th}_{\ps}$  variation during the run.
The $\vt_2$  and HCAL efficiency defined by  the leak of the signal shower energy from the ECAL to these detectors,   was studied  for different $X$ masses with simulations that were validated with measurements at the $e^- $ beam. The uncertainty in the efficiencies dominated mostly by the pileup effect was estimated to be $\lesssim 4\%$.  The trigger efficiency  was found to be $0.95\pm 0.02$. 
The $X$ signal-event acceptance  was estimated  by taking into account the efficiency of selection cuts for the signal shower shape in the ECAL \cite{gkkk}.  
The dominant uncertainty in the signal yield $\simeq 10\%$ was conservatively  accounted for  the difference between the predicted and measured dimuon yield \cite{na64prd18}. 
The total signal  efficiency $\epsilon_{X}$  was in the range  0.5 - 0.7  depending on the beam intensity and  the  $X$ mass. 
\par  
  To set the limits we analysed  runs I-III  simultaneously using the technique 
 based on the RooStats package \cite{root} allowing multibin limit setting \cite{na64prd18}. For  each of $X=S,~P,~V,~A$ cases,  we tried to optimize the size of  the signal box
 by  comparing sensitivities defined as an average expected limit calculated using the profile likelihood method. The calculations were done  
 by taking into account  the background estimate, efficiencies,  and their  corrections  with uncertainties used as nuisance parameters \cite{Gross:2007zz}. 
 For this optimization, the most important inputs came from the background extrapolation into the signal region from  the data 
samples of  runs I-III with their errors estimated from the extrapolation procedure. 
The optimal  signal box size  was found to be  weakly dependent on the  $e-X$ type  of  interaction and $X$   mass varying with a few GeV,  and was  finally set to $E_{\ec} \lesssim 50$ GeV
for  all  four cases of Eq.(\ref{eq:Lagrangian}) and the whole mass range.  
\begin{figure}[tbh]
\begin{center}
\vskip0.cm{\includegraphics[width=0.5\textwidth]{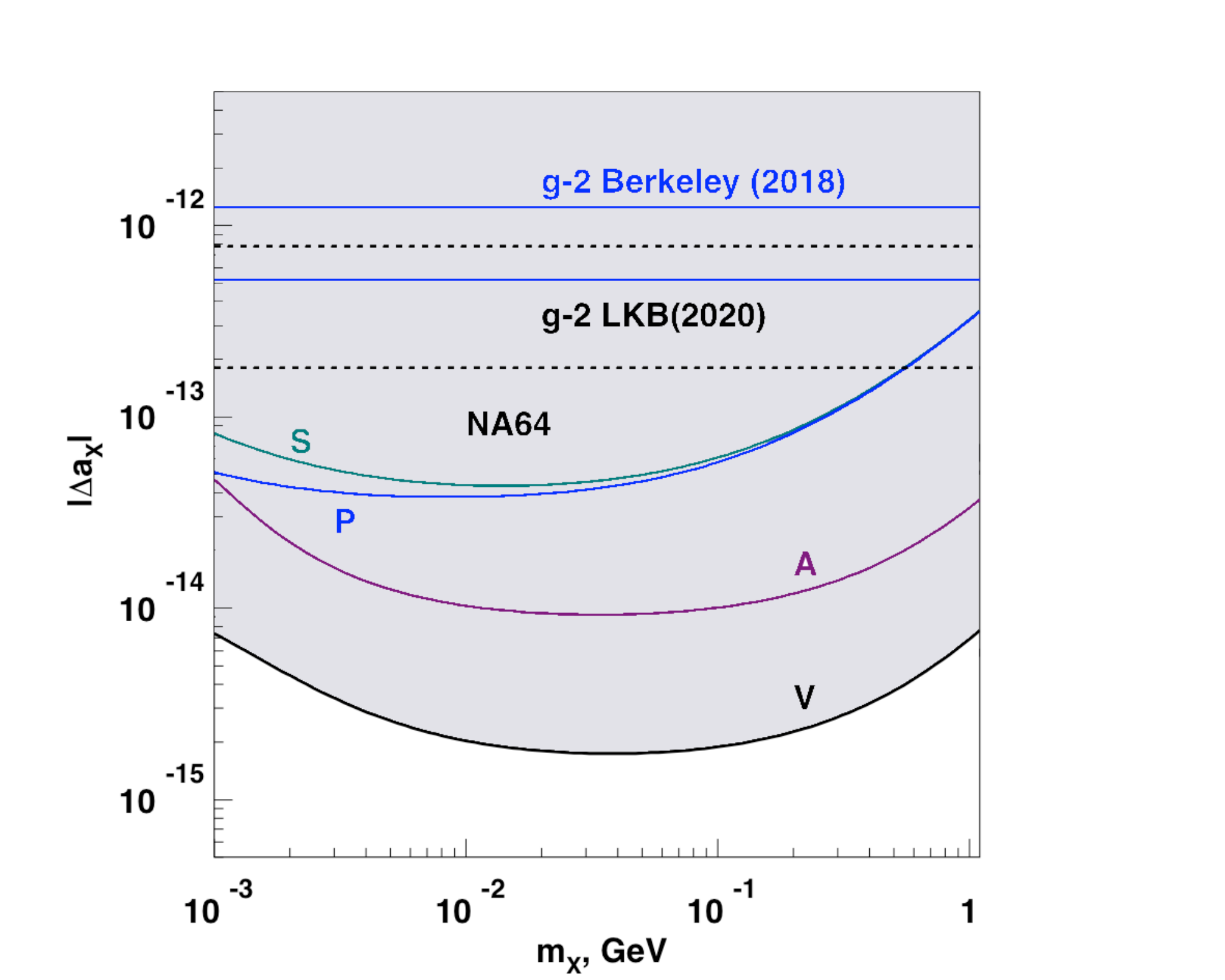}}
\vskip0.cm{\caption {Shown are the NA64 90\% C.L. exclusion region in the ($m_{X}, |\Delta a_X| $) 
plane for the $ S, P, V$ and $A$  contributions to $a_e$  together with the bands  of Eqs.(\ref{eq:lkb},\ref{eq:berk}),   representing the  results of the LKB \cite{lkb} (black dashed) and 
Berkeley \cite{berkeley} (blue solid)  experiments. The legend is the same as for Fig. 3. 
  \label{fig:delta-a}}}
\end{center}
\end{figure} 
The total number of signal events was the sum of expected events from the all three runs in the signal box:
\begin{equation}
N_X = \sum_{i=1}^{3} N_{X}^i = \sum_{i=1}^{3} n_{EOT}^i  \epsilon_{X}^i n_{X}^i (g_X, m_X, \Delta E_{e})
\label{nev}
\end{equation}
where $\epsilon_{X}^i$ and $n_{X}^i(\epsilon,\mx, \Delta E_{X})$ is the signal efficiency and  the 
signal yield per EOT in the energy  range $\Delta E_{e} $, respectively. These values were calculated from simulations  and  processing of signal events through 
the reconstruction program  with the same selection cuts and  efficiency corrections as for the  data sample from run $i$.
\par The combined 90\% C.L. exclusion limits on the coupling parameter $\varepsilon_X$   as a function of the $X$ mass,  calculated by using the modified frequentist approach  \cite{na64prl19, junk,limit,Read:2002hq} 
are shown in Fig.~\ref{fig:limits}. By using Eqs.(\ref{eq:lkb}), (\ref{eq:berk}) and (\ref{eq:contribS}) - (\ref{eq:contribA}),  it is also possible to translate the measurements of 
Refs.\cite{lkb,berkeley}  into constraints on the coupling $\varepsilon_X$ which are  shown in Fig. \ref{fig:limits} for comparison.  The limits were calculated by 
taking into account  the  sign of the contributions $\Delta a_X$  in  Eqs.(\ref{eq:contribS}) - (\ref{eq:contribA}) assuming that  the $S$ and $V$  contribute to the deviation  of Eq.(\ref{eq:lkb}) , while only  the $P$ and $A$  can resolve the discrepancy of  Eq.(\ref{eq:berk}). 
Our  bounds  are more stringent than those derived  from  the results of high-precision measurements  of Refs.\cite{gabri,lkb,berkeley}. Using 
  Eqs.(\ref{eq:contribS}) - (\ref{eq:contribA})  and obtained limits on the $X$ coupling  strength we can  derive  constraints on the $X$ contribution $\Delta a_X$ to  $a_e$. 
This results in stringent bounds in the range  $|\Delta a_X| \lesssim 10^{-15} - 10^{-13}$ for  {\it S, P, V} and {\it A} with sub-GeV masses,  which are shown 
 in the $(m_X; |\Delta a_X|)$   plane in Fig.~\ref{fig:delta-a}  together with the experimental bands of the $\Delta a_X$  values defined  by Eqs.(\ref{eq:lkb}, \ref{eq:berk}).
 For the low mass region $m_X\lesssim 10$ MeV  the limits were obtained by taking into account corrections from the exact calculations.
 These results are  an order of magnitude more sensitive compared to the current accuracy on $a_e$ from  the electron $g-2$ experiments  and recent high-precision determination of the fine structure constant, thus demonstrating  the strength of the NA64 approach on probing new physics in the electron $g-2$.
\par We gratefully acknowledge the support of the CERN management and staff 
and the technical staff of the participating institutions for their vital contributions.  
 This work was supported by the  Helmholtz-Institut f\"ur Strahlen-und Kern-physik (HISKP), University of Bonn, the Carl Zeiss Foundation 0653-2.8/581/2, and  Verbundprojekt-05A17VTA-CRESST-XENON (Germany), Joint Institute for Nuclear Research (JINR) (Dubna), the Ministry of Science and Higher Education (MSHE)  and RAS (Russia), ETH Zurich and SNSF Grant No. 197346, 186181 and 186158 (Switzerland), 
FONDECYT Grants No.1191103 and No. 1190845, SAPHIR Millennium Institute of ANID and ANID PIA/APOYO AFB180002 (Chile).


\begin{thebibliography}{99}

\bibitem{jr}
  J.~Jaeckel and A.~Ringwald,
   Annu.\ Rev.\ Nucl.\ Part.\ Sci.\  {\bf 60}, 405 (2010).

\bibitem{Essig:2013lka} 
  R.~Essig {\it et al.},  arXiv:1311.0029.


 \bibitem{report1}
  J.~Alexander {\it et al.}, arXiv:1608.08632.

\bibitem{report2} 
  M.~Battaglieri {\it et al.},  arXiv:1707.04591.


\bibitem{pbc-bsm}
J.  Beacham  {\it et al.},  J.\ Phys.\ G {\bf 47},  010501 (2020); arXiv:1901.09966.   


\bibitem{esu} R.~ K.~ Ellis at el. (European Strategy for Particle Physics Preparatory Group), arXiv:1910.11775.

\bibitem{berlin}
A. Berlin, N. Blinov, G. Krnjaic, P. Schuster, and N. Toro,
Phys. Rev. D {\bf 99}, 075001 (2019).

\bibitem{gaia} G.~Lanfranchi, M.~ Pospelov, and P.~Schuster, arXiv: 2011.02157.

\bibitem{bf} C.~Boehm and P.~Fayet, Nucl. Phys. {\bf B} 683, 219 (2004).


\bibitem{pf}
  P.~Fayet,
  Phys.\ Rev.\ D {\bf 75},   115017 (2007).


\bibitem{prv}
  M.~Pospelov, A.~Ritz, and M.~B.~Voloshin,
  Phys.\ Lett.\ B {\bf 662}, 53 (2008).

\bibitem{mp}
  M.~Pospelov, Phys. Rev. D {\bf 80}, 095002 (2009).

\bibitem{arkani} N.~Arkani-Hamed, D.~P.~ Finkbeiner, T. ~R.~ Slatyer, and N.~Weiner, Phys. Rev. D {\bf 79}, 015014 (2009).


\bibitem{g-2_mu} G. W. Bennett et al. (Muon g-2), Phys.\ Rev.\ D {\bf 73}, 072003 (2006).

\bibitem{atomki1}  A.~J.~Krasznahorkay,  et al., Phys. Rev. Lett. {\bf 116}, 042501 (2016). 
\bibitem{atomki2}  A.~J.~Krasznahorkay,  et al.,  arXiv:1910.10459. 

\bibitem{gk}
  S.~N.~Gninenko and N.~V.~Krasnikov,
  Phys.\ Lett.\ B {\bf 513}, 119 (2001).
  
\bibitem{feng} J. L. Feng, B. Fornal, I. Galon, S. Gardner, J. Smolinsky, T. M. P. Tait, and P. Tanedo, Phys. Rev. Lett. {\bf 117}, 071803 (2016).


 \bibitem{fnal} J.~Grange et al. (Muon g-2), arXiv:1501.06858.

\bibitem{jpark} T.~Mibe (J-PARC g-2), Chin. Phys. C {\bf 34}, 745 (2010).


\bibitem{na64be1} D.~Banerjee {\it et al.} (NA64 Collaboration),  Phys.\ Rev.\ Lett.\  {\bf 120}, 231802  (2018).
\bibitem{na64be2} D.~Banerjee {\it et al.} (NA64 Collaboration),  Phys.\ Rev.\ D {\bf 101},  071101  (2020).

\bibitem{na64be3} E.~Depero {\it et al.} (NA64 Collaboration),  Eur. Phys. J. C {\bf 80}, 1159 (2020).

 
\bibitem{lkb}
L.~Morel, Zh.~Yao, P.~Clad\'e,  and S.~Guellati-Kh\'elifa, Nature {\bf 588}, 61 (2020).

\bibitem{berkeley}  R.~H.~Parker, C.~Yu, W.~Zhong,  B.~Estey, and  H.~M\"uller,  Science {\bf 360}, 191 (2018).

\bibitem{aoyama1}  T.~Aoyama, M.~Hayakawa, T.~Kinoshita, and M.~Nio, 
Phys. Rev. Lett. {\bf 109}, 111807 (2012).

\bibitem{aoyama2}  T.~Aoyama, T.~Kinoshita, and M.~Nio,  Atoms {\bf 7}, 28 (2019).


\bibitem{gabri} D.~ Hanneke,  S.~ Fogwell, and G.~ Gabrielse, 
Phys.\ Rev.\ Lett.\ {\bf 100}, 120801 (2008).


\bibitem{giud} G.~F.~Giudice, P.~Paradisi, and M.~Passera, J. High Energy Phys.  {\bf11} (2012) 113.








\bibitem{nvk}
N.~V.~ Krasnikov, Mod.\ Phys.\ Lett.\ A {\bf 35}, 2050116 (2020).

\bibitem{Lee:2014tba}
  H.~S.~Lee,
  Phys.\ Rev.\ D {\bf 90},    091702(R)  (2014).


\bibitem{marci1} H.~Davoudiasl and W.~J.~Marciano, Phys.\ Rev.\ D {\bf 98},   075011 (2018).

\bibitem{marci2} W.~J.~Marciano,  A.~Masiero, P.~Paradisi, and M.~Passera,  Phys.\ Rev.\ D {\bf 94}, 115033 (2016).

\bibitem{criv} A.~Crivellin, M.~Hoferichter, and P.~ Schmidt-Wellenburg,     Phys.\ Rev.\ D {\bf 98},   113002 (2018).

\bibitem{chun} E.~J.~ Chun, T.~ Mondal,  J. High Energy Phys.  {\bf 11} (2020) 077.

\bibitem{liu} J.~Liu, C.~ E.~M.~ Wagner, and  X.-P.~ Wang,  J. High Energy Phys. {\bf 03},  (2019)  008.

\bibitem{bau}    M.~Bauer, M.~Neuber, S. Renner, M. Schnubel, and A.~ Thamm,  Phys.\ Rev.\ Lett. {\bf 124},  211803 (2020).

\bibitem{susi} S.~Gardner, X.~Yan, Phys.\ Rev.\ D {\bf 102}, 075016 (2020). 

\bibitem{darme} L.~Darme, F.~Giacchino, E.~Nardi, and M. Raggi, arXiv: 2012.07894.

\bibitem{rose} L.~D.~Rose, S.~Khalil, and S.~Moretti, arXiv: 2012.06911 [hep-ph].
 
\bibitem{endo} M.~Endo, W.~Yin,  J. High Energy Phys. {\bf 08}  (2019) 122.

\bibitem{ilja} I.~Dorsner,  S. Fajfer, and S. Saad,  Phys.\ Rev.\ D {\bf 102},  075007 (2020).

\bibitem{chen} K.-F.~Chen, C.-W.~ Chiang, and  K.~Yagyu, J. High Energy Phys. {\bf 09}  (2020) 119.
 
\bibitem{celine} C.~Boehm,  X.~Chu,  J.-L.~ Kuo, J. Pradler, arXiv: 2010.02954. 

\bibitem{gkm} S.~N.~Gninenko, N.~V.~Krasnikov,  and V.~A.~Matveev,   Phys.\ Part.\ Nucl.\  {\bf 51},  829 (2020).

\bibitem{pdg} 
P.~A.~Zyla et al. (Particle Data Group), Prog. Theor. Exp. Phys. {\bf 2020}, 083C01 (2020).

 \bibitem{na64prl19} 
  D.~Banerjee {\it et al.} (NA64 Collaboration),
  Phys.\ Rev.\ Lett.\  {\bf 123},  121801 (2019).

\bibitem{babarg-2} 
  J.~P.~Lees {\it et al.} ({\it BABAR} Collaboration), Phys. Rev. Lett. {\bf 119}, 131804 (2017).
  
  \bibitem{Gninenko:2018ter}
  S.~N.~Gninenko, D.~V.~Kirpichnikov, and N.~V.~Krasnikov,  Phys.\ Rev.\ D {\bf 100}, 035003 (2019).
 
  
\bibitem{Kirpichnikov:2020tcf}
  D.~V.~Kirpichnikov, V.~E.~Lyubovitskij, and A.~S.~Zhevlakov,
  Phys.\ Rev.\ D {\bf 102},  095024 (2020).  
  
\bibitem{Gninenko:2013rka}
S.~N.~Gninenko,
  Phys.\ Rev.\ D {\bf 89},  075008 (2014).
  

\bibitem{Andreas:2013lya} 
  S.~Andreas {\it et al.},  arXiv:1312.3309.
  
  
\bibitem{Banerjee:2015eno}
  D.~Banerjee, P.~Crivelli, and A.~Rubbia,
   Adv.\ High Energy Phys.\  {\bf 2015},  105730 (2015).


\bibitem{na64srd}  
  E.~Depero {\it et al.},
  Nucl.\ Instrum.\ Methods\ Phys. Res., Sect. A {\bf 866},  196 (2017).


 \bibitem{na64prl17} 
  D.~Banerjee {\it et al.} (NA64 Collaboration),
  Phys.\ Rev.\ Lett.\  {\bf 118},   011802 (2017).
 

\bibitem{na64prd18} 
  D.~Banerjee {\it et al.} (NA64 Collaboration),
  Phys.\ Rev.\ D {\bf 97},  072002 (2018).

\bibitem{Agostinelli:2002hh}
  S.~Agostinelli {\it et al.} [GEANT4 Collaboration],
  Nucl.\ Instrum.\ Methods\ Phys. Res., Sect. A {\bf 506}, 250  (2003).

\bibitem{geant} 
  J.~Allison {\it et al.},
  IEEE Trans.\ Nucl.\ Sci.\  {\bf 53}, 270 (2006).
\bibitem{dmg4} 
A. Celentano, M. Bondi, R. R. Dusaev,  D. V. Kirpichnikov, M. M. Kirsanov, N. V. Krasnikov, L. Marsicano, and D. Shchukin, arXiv: 2101.12192.




 

\bibitem{gkkk} 
  S.~N.~Gninenko, N.~V.~Krasnikov, M.~M.~Kirsanov, and D.~V.~Kirpichnikov,
  Phys.\ Rev.\ D {\bf 94},  095025 (2016).
  
 \bibitem{gkkketl}
  S.~N.~Gninenko, D.~V.~Kirpichnikov, M.~M.~Kirsanov, and N.~V.~Krasnikov,   
   Phys. Lett. B {\bf 782}, 406 (2018).

\bibitem{dkk} 
R.~R.~ Dusaev, D.~V.~ Kirpichnikov, and M.~M.~ Kirsanov, Phys.\ Rev.\ D {\bf102}, 055018 (2020). 




\bibitem{root} 
  I.~Antcheva {\it et al.},
  Comput.\ Phys.\ Commun.\  {\bf 180}, 2499 (2009).

\bibitem{Gross:2007zz}
  E.~Gross,
  ``LHC statistics for pedestrians,'' CERN Report No. CERN-2008-001, 2008, p.71.



\bibitem{junk}
T. Junk, 
Nucl. Instrum. Methods Phys. Res., Sect. A {\bf 434}, 435 (1999).


\bibitem{limit}
G. Cowan, K. Cranmer, E. Gross, and O. Vitells, 
Eur. Phys. J. C {\bf 71}, 1  (2011).

\bibitem{Read:2002hq}
  A.~L.~Read,
  J.\ Phys.\ G {\bf 28}, 2693 (2002).

  
\end{thebibliography}
\end{document}